\documentclass[12pt,preprint]{aastex} 
\slugcomment{Submitted to The Astrophysical Journal} 

\begin{document} 
\title{HST/WFPC2 Images of the GG Tauri Circumbinary Disk} 
\author{John E. Krist\altaffilmark{1}, 
Karl R. Stapelfeldt\altaffilmark{2},
and Alan M. Watson\altaffilmark{3}} 
\altaffiltext{1}{Space Telescope Science Institute, 3700 San Martin Dr., 
Baltimore, MD 21218 (krist@stsci.edu)}
\altaffiltext{2}{MS 183-900, Jet Propulsion Laboratory, Pasadena, CA 91109
(krs@exoplanet.jpl.nasa.gov)}
\altaffiltext{3}{Instituto de Astronomia, Universidad Nacional Aut\'onoma
de M\'exico, Apartado Postal 72-3, 58089 Morelia, Michoac\'an, M\'exico
(alan@astrosmo.unam.mx)}

\begin{abstract} 
We present the first visible wavelength images of the GG Tauri
circumbinary ring, obtained with the Hubble Space Telescope's Wide Field
and Planetary Camera 2.  Scattered light from the ring is detected in
both V and I band images.  The images show that the ring is smooth,
except for a small gap that could be a shadow caused by material between
the stars and the ring.  The spokes seen extending from the stars to the
ring in ground-based adaptive optics images are not seen in our data, 
which suggests that they may be image artifacts.  The nearside/farside 
surface brightness ratio is 6.9 in I band, consistent with forward
scattering by small dust grains.  The azimuth of the peak
ring surface brightness appears offset by 13$^{\circ}$ from the
azimuth closest to us, as seen in previous near-IR HST observations.  This
may indicate that the ring is warped or somehow shadowed by the circumstellar
disks.  The color of the ring is redder than the
combined light from the stars as observed by HST, confirming previous
measurements that indicate that circumstellar disks may introduce extinction
of light illuminating the ring.  We detect a bright, compact arc of material 
0\farcs 3 from the secondary star at an azimuth opposite the primary.  It
appears to be too large to be a circumstellar disk and is not at the
expected location for dust trapped at a Lagrange point.
\end{abstract}


\section{Introduction}

GG Tauri (HBC 54; IRAS 04296+1725) is a young multiple star system
located in the Taurus L1551 molecular cloud at a distance of 140 pc
(Elias 1978).  It includes two binaries, GG Tau {\it Aa/Ab}, with a 
separation of $\sim$0\farcs 25, and 10\farcs 6 away, GG Tau {\it Ba/Bb}, 
separated by 1\farcs 5.  GG Tau {\it Aa/Ab} (collectively hereafter 
just GG Tau) is especially interesting because it possesses a circumbinary 
disk that has been imaged in molecular line and dust continuum emission 
with millimeter interferometry (Dutrey, Guilloteau, \& Simon 1994, hereafter 
DGS94; Guilloteau, Dutrey, \& Simon 1999, hereafter GDS99).  Subsequently, 
images of the disk in scattered light were obtained in the near-infrared with
ground-based adaptive optics (Roddier et al. 1996, hereafter R96) and the
Hubble Space Telescope (HST) (Silber et al. 2000; McCabe \& Ghez 2000).

These measurements show that the disk has a large inner hole, $\sim$180
AU in radius, which has apparently been cleared by tidal interactions
with the binary.  While the disk has been detected out to a radius
of $\sim$800 AU in millimeter line emission, virtually all of the
scattered light and millimeter continuum emission is confined within
an annulus of radius 180-260 AU.  For this reason, the disk is usually
described as a ring.  The morphology is consistent with a circular
ring inclined $\sim$37$^{\circ}$ from the plane of the sky.  The CO
kinematics are in excellent agreement with Keplerian disk models (GDS99).
Redshifted gas is found along the ring's western side, and blueshifted
gas on its eastern.  Assuming that the disk and binary rotate in the same
direction, this information and the observed motions of the stars indicate
that the northern, brighter side of the ring must be the nearest to us
(R96).  

The adaptive optics (AO) images of R96, which were further analyzed by
Close et al. (1998), showed a clumpy ring and indicated the presence
of radial ``spokes'' extending inward from the ring toward the central
binary.  Later AO images taken with Gemini (Potter et al. 2001) also contained
some spokes.  These features resembled predictions for accretion streams from
a circumbinary disk (Artymowicz \& Lubow 1994).  However, both of the
near-IR observations with HST (Silber et al. 2000; McCabe \& Ghez 2000) show 
an essentially smooth ring,
with no evidence for the spokes at the intensity levels indicated in
the AO images.

Previous photometry of the individual binary components suggests that
each may possess its own circumstellar disk.  Near-infrared excesses
were noted in both stars by R96.  White et al. (1999) derived from HST
spectra line-of-sight extinctions A$_V$=0.72 and 3.20 to GG Tau {\it Aa} and
{\it Ab} respectively.  This large difference between two sources just 
0\farcs 25 (35 AU) apart (projected) shows that there must be extinction localized at
one of the stars, perhaps from a small circumstellar disk.  Limits to the
amount of circumstellar material are provided by high resolution 2.7 mm
and 1.3 mm continuum maps by Looney, Mundy, \& Welch (2000) and GDS99,
respectively.  These show that no more than a few percent of the total
flux from the system can originate in the circumstellar material of
the individual binary components.  While any such disks must therefore
be small and low mass in comparison to the outer circumbinary ring,
their presence may have important effects on the ring's illumination.
Indeed, R96 noted that the ring was redder than the combined light from
the two stars and suggested that this indicated extinction along the line
of sight between the central binary and the ring.  Wood, Crosas, \& Ghez
(1999) explored this scenario quantitatively, and through modeling of the
scattered light images of R96 and $^{13}$CO J=1-0 line profiles of DGS94,
found that extinction from circumstellar disk(s) appeared necessary to
account for the brightness and color of scattered light from the ring.
Despite the many lines of indirect evidence, no resolved images of
circumstellar disks in the GG Tau binary have yet been obtained.

Based on its strong IR excess and millimeter continuum flux, GG Tau
was included in the Wide Field and Planetary Camera 2 (WFPC2) Investigation
Definition Team's HST observing program before the nature of the ring
was confirmed by DGS94 and R96.  We present these results here, the first 
images of the GG Tau ring at visual wavelengths.

\section{Observations and Processing}

GG Tau was observed using HST/WFPC2 on 27 September 1997 as part of HST
program 6855.  High resolution images of the binary and ring were taken
at the center of the Planetary Camera (PC1, 0\farcs 0456 pixel$^{-1}$)
in F555W (WFPC2 V band) and F814W (WFPC2 I band).  Short exposures (3s
in F555W, 2s in F814W) provided unsaturated images of the binary for
photometry, while longer exposures (two 200s in F555W and two 100s in
F814W) were used for imaging the ring.  In addition, short (1s), medium
(40s) and long (two 300s) exposures were taken through filter F606W with
GG Tau centered in Wide Field Camera 3 (WF3, 0\farcs 1 pixel$^{-1}$),
in order to maximize sensitivity to low surface brightness emission.
All images were taken at a gain of 14 photons DN$^{-1}$, except for 
the longest exposure F606W images which were taken at 7 photons DN$^{-1}$.

The data were processed by the HST calibration pipeline, and duplicate
images combined with cosmic ray rejection.  In the longer exposures, the
stars were saturated and there was CCD column bleeding.  These saturated
pixels were replaced by scaled, unsaturated values from shorter exposures.
This produced images that were unsaturated beyond 0\farcs 14.

\begin{deluxetable}{lrrr}
\tabletypesize{\scriptsize}
\tablecaption{Photometry of GG Tauri Stars and Disk Features. \label{tbl-1}}
\tablewidth{0pt}
\tablehead{
\colhead{} & \colhead{V} & \colhead{I$_c$} & \colhead{V-I$_c$}\\ 
\colhead{Object} & \colhead{(mag)} & \colhead{(mag)} & \colhead{(mag)}
}
\startdata
GG Tau Aa & 12.25$\pm$0.03 & 10.47$\pm$0.03 & 1.78 \\
GG Tau Ab & 14.70$\pm$0.06 & 11.98$\pm$0.04 & 2.72 \\
GG Tau Ba & 17.11$\pm$0.07 & 13.56$\pm$0.06 & 3.55 \\
GG Tau Bb & 19.94$\pm$0.08 & 15.55$\pm$0.07 & 4.39 \\
Ring (PA=20$^{\circ}$)\tablenotemark{1} & 18.40$\pm$0.08 & 15.57$\pm$0.09 & 2.83 \\
Ring (PA=142$^{\circ}$)\tablenotemark{1} & $>$20.29$\pm$0.50 & 17.75$\pm$0.20 & $>$2.54 \\
Ring (PA=187$^{\circ}$)\tablenotemark{1} & $>$20.10$\pm$0.50 & 17.35$\pm$0.20 & $>$2.75 \\
\enddata
\tablenotetext{1}{Values are mag arcsec$^{-2}$.}
\end{deluxetable}

\subsection{Point Spread Function (PSF) Photometry and Astrometry}

Using a method that has been effective in the past for close binaries
such as FS Tauri (Krist et al. 1998) and XZ Tauri (Krist et al. 1999),
the position and intensity of each component of GG Tau was determined 
by simultaneously fitting simulated PSFs within 39 by 39 pixel boxes centered
on the binary in the short exposures.  The PSFs were generated for each
filter using version 5.0c of the Tiny Tim HST PSF modeling program
(Krist \& Hook 2000).  The models matched the reported colors (Ghez,
White, \& Simon 1997) of each binary component.  They were subsampled
by a factor of five in each dimension, allowing the fitting procedure to
accurately interpolate and rebin the PSFs to match the positions of the
stars in the images.  Aperture corrections derived from larger PSF models 
were applied to the measured
fluxes which were then converted to V and I$_c$ magnitudes using SYNPHOT
(Bushouse \& Simon 1998), assuming an M1.5V spectral type.  These values
are provided in Table 1.  The uncertainties include estimates of 
the measurement errors and errors in the conversion of fluxes to standard
magnitudes.  The combined magnitudes are V=12.15 and I$_c$=10.23 
(V-I$_c$=1.92).  

The measured separations are 0\farcs 251 (F814W) and 0\farcs 249 (F555W),
with estimated errors of $\pm$0\farcs 003.  The average position angle
is $354.3^{\circ}\pm1^{\circ}$.  In the 3 years between our data and the
September 1994 WFPC2 images of Ghez, White, \& Simon (1997), the position 
angle has decreased by $\sim$4.5$^{\circ} \pm 2^{\circ}$ ($\sim$1.5$^{\circ}
\pm 0.7^{\circ}$ year$^{-1}$) but the separation has not measurably changed.
This agrees with the astrometric results compiled by Woitas, K\"ohler, \&
Leinert (2001).  The combined mass of GG Tau {\it Aa} and {\it Ab} is 
$1.28 \pm0.07$ M$_{\odot}$, as inferred from the ring kinematics by Simon, 
Dutrey, \& Guilloteau (2000).  The mean angular velocity for a circular binary 
orbit with this mass and semi-major axis of 18 AU (22 AU deprojected; d=140 pc)
is 4.0$^{\circ}$ (2.8$^{\circ}$) per year.  The observed rate is much less 
than this, so the orbit is likely to be eccentric (e$>0.25$) and observed
near apoastron (assuming that the binary and ring are coplanar).

In addition to the PSF-fitting photometry of {\it Aa}/{\it Ab}, we also used
aperture photometry to measure the fluxes of the GG Tau {\it Ba}/{\it Bb} stars,
which are shown in Table 1.  Conversions to standard magnitudes were done
using SYNPHOT and an M5.5V spectrum.  GG Tauri {\it Bb} is especially red 
compared to the other stars and has been identified as a possible young brown 
dwarf (White et al. 1999).

\begin{figure}
\plotone{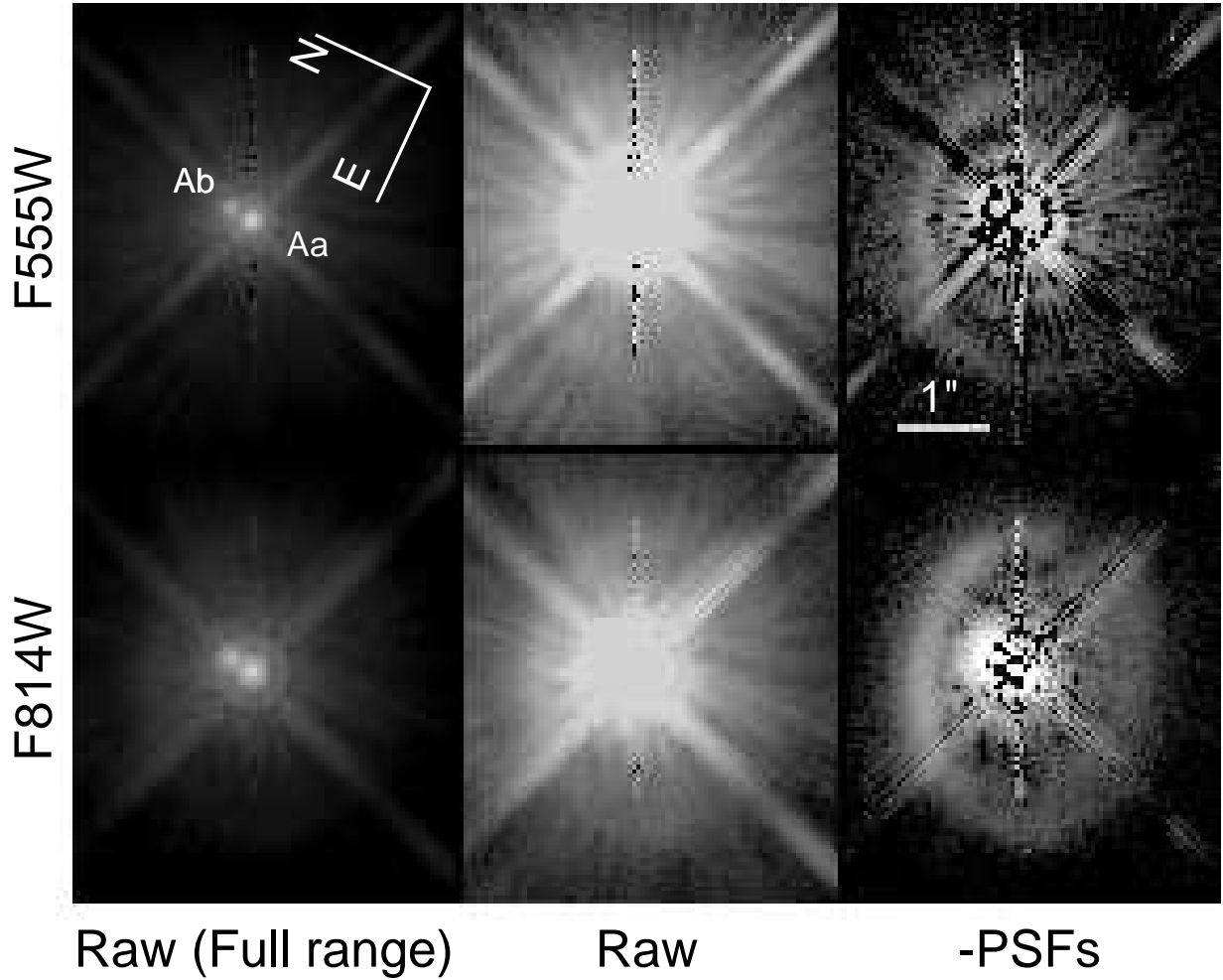}
\caption{
HST/WFPC2 Planetary Camera (0\farcs 0456 pixel$^{-1}$) images of GG
Tauri before and after PSF subtraction (logarithmic stretches).  On
the left are the images displayed over their full range.  The 
middle and right frames are identically stretched to emphasized
low-brightness features.}
\end{figure}

\subsection{PSF Subtraction}

The ring intensity in the long exposures was just above or equal to the
local background level, which was dominated by the wings of the combined
point spread functions (Figure 1).  PSF subtraction was required to provide
a clear
view of the ring.  This is complicated by the optical characteristics
of HST and WFPC2, which cause the PSF to be dependent on position,
focus, and color, introducing potential mismatches between target and
reference images.  The best subtractions are obtained using images of
isolated stars that have similar colors and exposure levels, and which
are imaged near (within 1'') the same detector location as the target.
At radii beyond $\sim$0\farcs 3, the model PSFs produced by Tiny Tim
do not match as well as observed PSFs.  Therefore, we selected a few 
isolated star images from other HST programs that matched the GG Tau 
observations in color, exposure level, and position on the detector.

All of the reference PSFs were saturated, with no short exposures to
provide direct measurements of their intensities.  We thus resorted
to combining aperture photometry (excluding the saturated pixels)
with assumed electron well-depths for saturated pixels to derive fluxes
(Gilliland 1994).  These values were then used to normalize the reference
PSFs to match the fluxes of the GG Tau components in the long exposures.
A binary PSF reference image was created by taking the reference PSF and
coadding it with a version of itself that was offset and scaled to match
the two components of GG Tau.  This was then aligned with the observed
image by eye using cubic convolution interpolation.  The registrations
appeared to be accurate to within 0.05 pixels.

In F814W, there was large selection of archived PSFs from which to choose.
Two stars provided good matches to GG Tau in terms of field position and
color : DQ Tauri and HD 181204.  Both have M0V spectral types and are
located within 0\farcs 8 of GG Tau's detector coordinates.  DQ Tau is a 
spectroscopic binary with a sub-milliarcsecond separation, so it appears
as an unresolved system on PC1.  The image of HD 181204 was more 
deeply exposed than DQ Tau's, with saturated pixels extending 0\farcs 25 from
the core.  Saturated pixels in this frame were replaced by those
from an image of DQ Tau that was appropriately shifted and intensity-scaled.
This produced a combined reference PSF that had high signal in the wings but 
had only a small number of saturated pixels in the core, making it suitable 
for use to within about 0\farcs 1 of the GG Tau stars.

\begin{figure}
\plotone{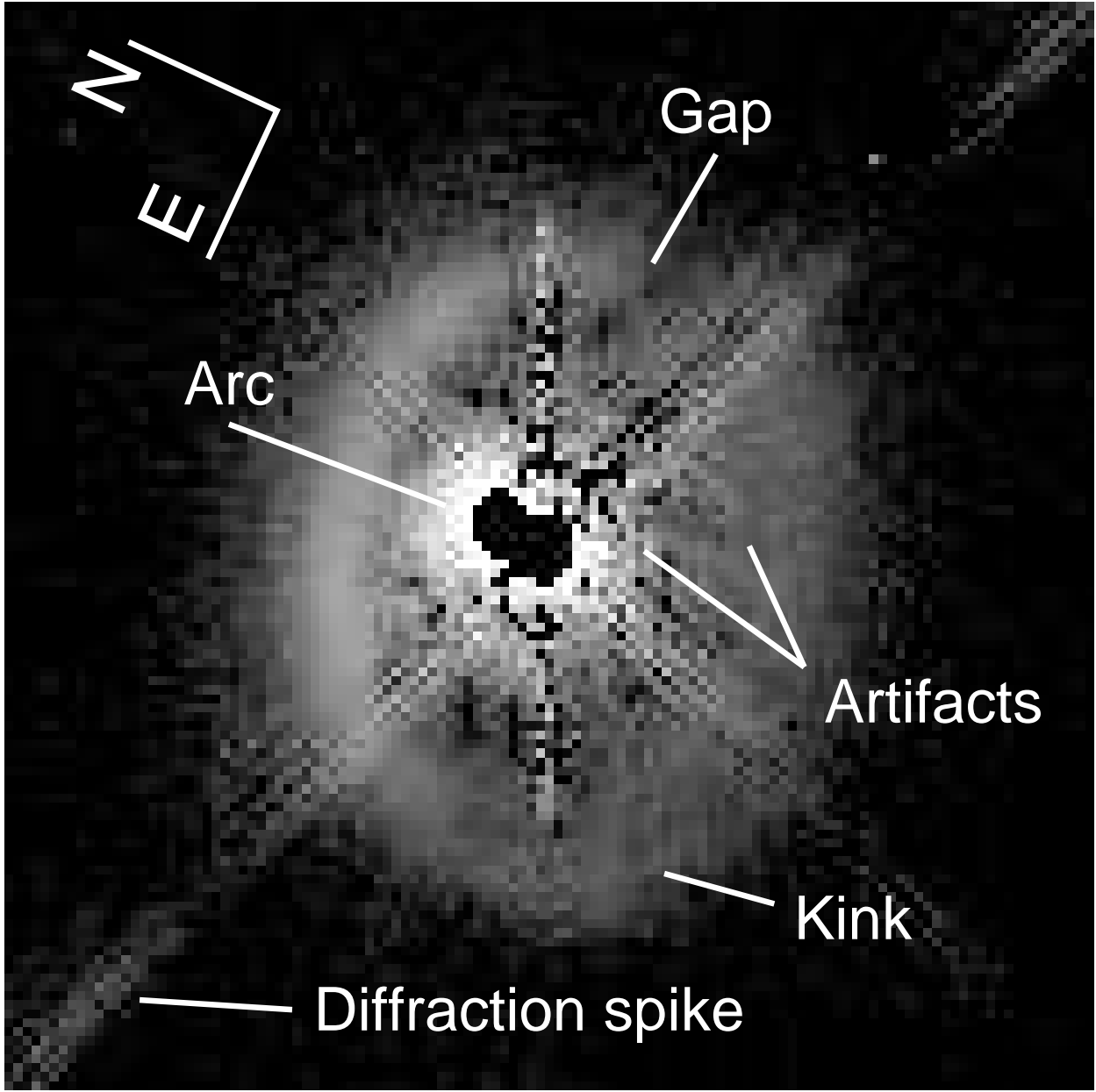}
\caption{
PSF-subtracted F814W image of the GG Tau ring, with various features
identified.  The orientation is the same as in Figure 1.  Saturated
pixels in the core of GG Tau have been masked.}
\end{figure}

The only good match to GG Tau in F555W was an image of LkCa 15, a
K5V star whose position on the detector was 34.4 pixels (1\farcs 6)
from GG Tau's.  LkCa 15 has a circumstellar disk which is seen from low
latitudes, and which was recently detected in scattered light using 
deep, dual-channel polarization, adaptive optics images using the Gemini 
telescope (Potter et al. 2001).  We failed to detect the LkCa 15 disk 
in this and other filters (F814W and F675W), likely due to the short
exposures used.  Given that the LkCa 15 scattered light disk is 
small (diameter of $\sim$1'', less than the central hole in the 
GG Tau ring) and within the level of the noise, it seemed relatively 
safe to use it for subtraction, and it was the only reasonable match.

The F814W subtractions produced good images of the ring, which is reliably
seen at all azimuths, excluding the diffraction spike locations.  The ring
interior is also largely free of artifacts. There is a small PSF residual 
ring around the brighter star (identified as an artifact in Figure 2).  The 
circumbinary disk in F814W is much better exposed than in F555W, due to the 
greater brightnesses of the stars in the longer wavelength passband.
The estimated detection limit (3$\sigma$ above the background) at 1'' is 
I$_c$=18.5 $\pm$ 0.25 mag arcsec$^{-2}$, based on an analysis of a large number 
of PSF subtractions that will be presented in a future paper (Stapelfeldt,
private communication).

The F555W subtraction is strongly contaminated by residuals due to mismatches
in both the colors and field positions of the target and reference PSFs.  The 
northern edge of the ring is sufficiently bright to dominate over the residuals,
except in the regions of diffraction spikes.  The southern edge, however, is 
clearly dominated by residual streaks and ringing patterns characteristic of PSF
mismatches, and so any detection of the ring in that region is suspect.
The interior of the ring in this filter is likewise filled with residuals,
none of which should be mistaken for physical structures.
We were unable to derive a good estimate of the detection limit for
this image due to the large PSF-related residuals but assume that it must be
brighter than V=20 mag arcsec$^{-2}$, based on the residual brightness in
the far side of the ring.  We strongly
urge observers with future high-contrast programs on WFPC2 to avoid our
problems and obtain a reference PSF of the appropriate color at a number
of positions around ($\sim$0\farcs 5 away from) the planned target
position, which will allow them to select the best match in terms of
location.

Unfortunately, we could not find in the HST archive an observed PSF for the 
WF3 F606W subtraction that suitably matched the color, field position, and 
exposure level of GG Tau.  Therefore, only the F555W and F814W Planetary Camera
images are discussed further.

\begin{figure}
\plotone{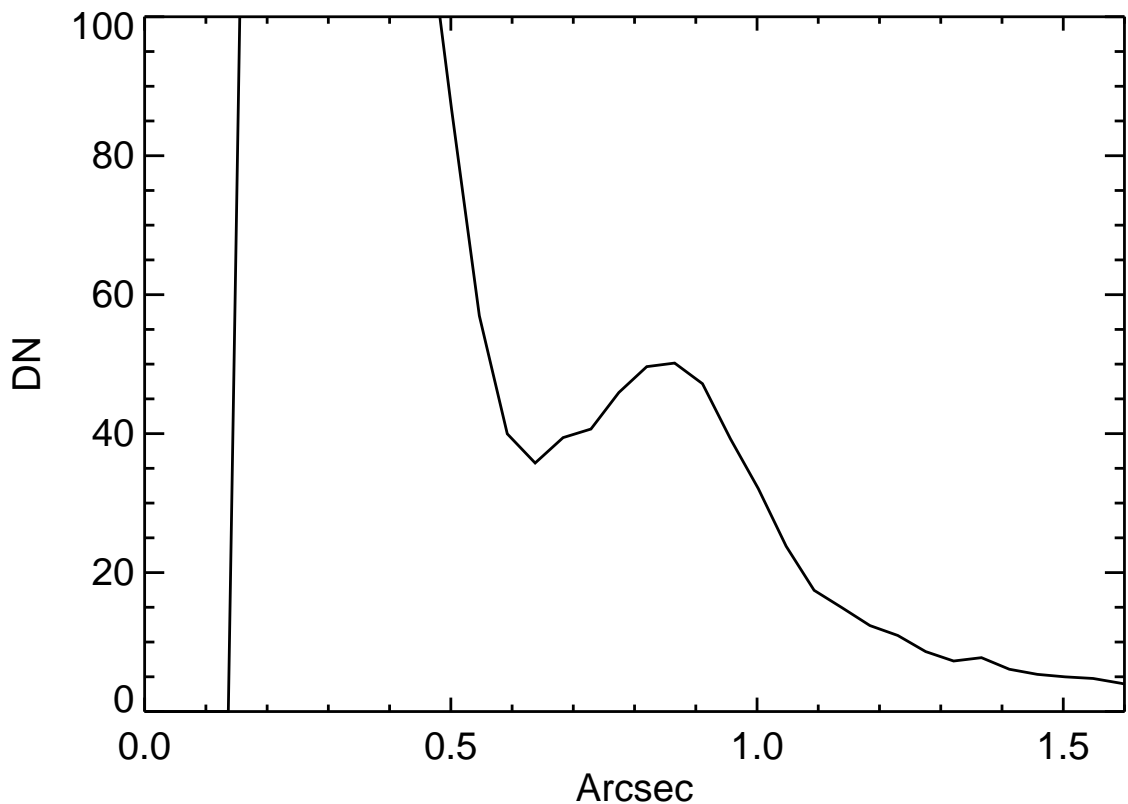}
\caption{
Radial intensity cut through the PSF subtracted F814W image of the circumbinary ring.
The cut is taken along an axis passing through the brightest part of the 
ring and spans position angles 14-22$^{\circ}$.  The linear distance and 
PAs are expressed relative to the midpoint between the two stars, and 
the amplitude is given in detector counts (proportional to photons).
The ring corresponds to the peak at r$=$0.85 arcsec; the relatively
broad profile is apparent.  Emission at r$<0.5$ arcsec traces the arc
near the secondary, and PSF residuals.  Saturated pixels at r$<0.15$ 
arcsec have been blanked out.
}
\end{figure}

\section{Results}

The raw and PSF-subtracted data are shown in Figure 1.
The PSF-subtracted images reveal a generally smooth ring, with no
prominent clumps.
A major feature of the ring is the broad azimuthal brightness peak
along its northern edge.  As noted by Silber et al.  2000, this is
almost certainly due to the forward scattering properties of small dust
particles within the ring.  Its radial intensity cross-section there is
approximately Gaussian, with a FWHM of 0\farcs 3 (Figure 3).
The surface brightness decreases by nearly a factor of two between
the brightest portion of the northern section of the ring inwards
to the darkest region of the clearing, a span of 0\farcs 15 (20 AU).
The contrast is probably larger than this, as PSF residuals may
dominate the surface brightness within the clearing.

The southern section of the ring appears about twice as broad as the
northern.  This is expected as we are seeing both the illuminated inner
wall and upper surface of the ring towards the south, as discussed by R96
and Silber et al. (2000).  The eastern extreme of the ring appears to have
a broader radial distribution than the western side, which is more narrowly 
peaked.  There is a slightly dark, azimuthal band in the southern 
side in F814W at {\it r}=1\farcs 1 that corresponds to a diffraction feature 
seen in the reference PSFs and is not real (i.e. it is likely not a shadow 
cast by a disk around one of the stars).  It is identified as an artifact 
in Figure 2.

There is a $\sim$0\farcs 4 wide gap in the eastern section
of the ring centered at PA$\approx268^{\circ}$ (Figure 2).  Its apparent 
southern edge is adjacent to a PSF diffraction spike artifact, so its full 
width cannot be determined from our images.  It is seen both in F555W and
F814W and is also apparent in the HST images of Silber et al. (2000)
and adaptive optics images (Potter et al.  2001; Tamura et al. 2001).
The brightness decreases by at least a factor of four within the
gap relative to the expected trend across this section of the ring.
At first glance it appears to be aligned towards the secondary star,
GG Tau {\it Ab}, but this could simply be a projection effect as we are seeing
both the inner wall and upper surface of the ring at that position.

\begin{figure}
\plotone{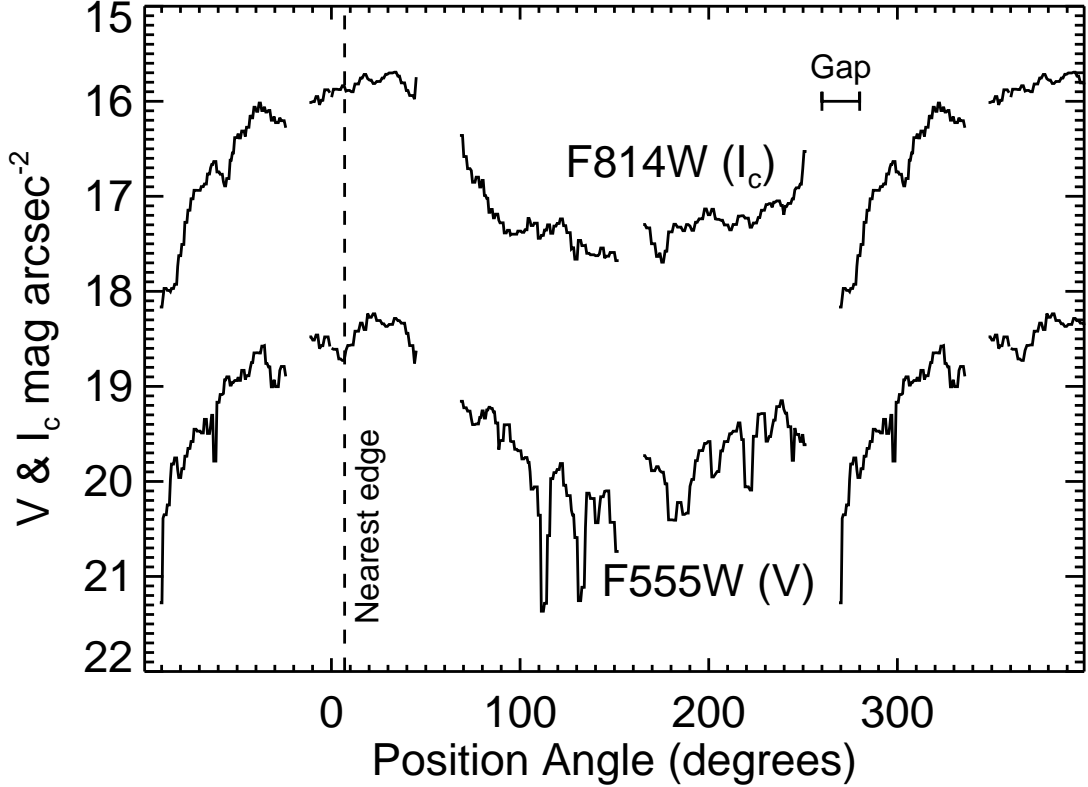}
\caption{
Azimuthal profile plot of the GG Tau ring.  The values plotted
are the mean surface brightness in a 3 by 3 pixel (0\farcs 14 by 0\farcs 14) 
aperture along an ellipse visually constrained to fit the ring.  Regions of no
data are the locations of the diffraction spikes.  The F555W plot has
a number of sharp local minima which are subtraction artifacts.  The position angle is
measured relative to the midpoint between the two stars.  Note that the
surface brightness of a PSF with the combined flux of both stars light at 1'' 
is V=17.5 and I=15.3 mag arcsec$^{-2}$.  The profiles have been replicated to
extend the plots beyond 0$^{\circ}$-360$^{\circ}$.  The position angle of
the nearest edge of the ring as determined from radio emission measurements
is marked with a dashed line.  The mean color difference is V-I$_c$=2.7.}
\end{figure}

Plots of the azimuthal ring surface brightness profiles are given in
Figure 4.  They were derived by computing the mean intensity in 3 by 3
pixel boxes along an ellipse that was manually constrained to pass through
the region of peak radial brightness at each azimuth around the ring.
These values were converted to V and I$_c$ surface brightnesses in the same
way as the stellar fluxes.  The profile shapes generally match between 
the two filters, though the surface brightness in F555W
is lower.  It is unclear if the F555W brightness in the southern part
of the ring is a secure detection - it is probably artificially high
there due to the PSF subtraction residuals.  Except for this region,
the ring has an overall V-I color of 2.7 mag, 0.8 mag redder than the
combined stellar flux.  The gap is evident in these plots
between PA=265-280$^{\circ}$, as marked in Figure 4.

The general peak intensity occurs at PA$\approx$25$^{\circ}$, in agreement with
Silber et al. (2000).  This does not coincide with the nearest edge of
the ring, which is at PA=7$^{\circ}$ as determined from mm observations
(GDS99).  The minimum lies between
PAs of 130-170$^{\circ}$, in a portion of the ring where the azimuthal
intensity trend is disrupted (a region that Silber et al. identified as
a ``kink''; see Figure 2).  The maximum and minimum surface brightnesses 
around the ring have a ratio of 6.9 in F814W, greater than the value of 
$\sim$4.0 at 1.2 $\mu$m measured by Silber et al., or the 5.0 at 1.6 $\mu$m
from McCabe \& Ghez 2000).  This ratio is expected to increase with
decreasing wavelength as forward scattering becomes stronger, assuming
that the bulk of the dust consists of $<$ 0.6 $\mu$m particles.

The interior of the ring, at least in the relatively clean F814W
subtraction, is brighter than the background outside of the ring, and it
appears clear of any physical structures except for a small arc (discussed 
below).  The interior surface brightness 
is I$_c <$ 18 mag arcsec$^{-2}$ (the large residuals in F555W are
too great to deduce an accurate interior brightness).  It is not clear
whether this flux is due to subtraction residuals or scattering material.
This amount of flux cannot, however, be explained by the convolution of 
the instrumental PSF with the ring.  We do not detect the bright artifact
Silber et al. saw in their subtraction at PA=144$^{\circ}$ midway between
the stars and the ring.  Unless it was a transient feature, we should
have seen it, as it was brighter than the nearby section of the ring. 
Two circular rings appear around the stars in both filters, and are
definitely subtraction artifacts.  There are no indications of bright 
spokes interior to the ring like those seen in the AO images of R96 
(though one of their spokes would lie under the region in our images 
contaminated by saturated column residuals).

In the F814W image there is a bright, compact structure adjacent to GG
Tau {\it Ab} that extends $\sim$0\farcs 3 towards the north from the star.
It has an arc shape (see Fig. 2), and is roughly symmetric about 
PA$\approx$0$^{\circ}$.  Its surface brightness decreases from I$_c$=12.7 to 16.5 
mag arcsec ($\pm$0.8 mag) between 0\farcs 2-0\farcs 6.
Within 0\farcs 2, the stellar subtraction residuals are too great to
obtain a good measurement, and residuals may dominate out to approximately
0\farcs 4.  The circumbinary ring dominates past 0\farcs 6.  A similar arc 
can be seen in the Silber et al. (2000) NICMOS images, though they do not 
point it out and presumably believed it to be a subtraction artifact.  It is 
not seen in the NICMOS images of McCabe \& Ghez (2000), which have significant
subtraction residuals near the stars and lower resolution.  Likewise,
we are unable to positively identify it relative to the large residuals
in our F555W image.  To investigate the potential that this arc could be
a subtraction residual like the ring seen around the primary in Figure 2, we 
generated a number of Tiny Tim model F814W PSFs
for a reasonable selection of object colors (K5V-M3V, including extinction), 
telescope focus offsets (0-6 $\mu$m breathing), and normalization errors 
(1-20\%), and then subtracted each from the others.
We were unable to duplicate this feature in form or relative brightness.
Likewise, the Silber et al. NICMOS images should not be biased by similar
effects.  These experiments, and the fact that the same structure is seen 
in the NICMOS data, lead us to conclude that it is actually reflecting 
material near GG Tau {\it Ab}.

\section{Discussion}

\subsection{Constraints on the current ring model}

The prevailing model for the GG Tau circumbinary disk is a vertically
thick, inclined circular ring with a completely cleared central hole and
illuminated by reddened starlight.  The WFPC2 images, as the highest
spatial resolution images yet obtained of the ring, and as the first
taken at wavelengths below 1 $\mu$m, provide new tests for this general
picture.  We now discuss the implications of the WFPC2 data
for the prevailing ring model.

\subsection{Ring geometry}

The isophotes in the F814W image were visually fitted with ellipses to
characterize the {\it apparent} ring geometry.
One focus of an ellipse was fixed at a location 46\% of the way from the
primary to the secondary star, in accordance with the inferred stellar
mass ratio (White et al. 1999).  The best-fit parameters were: inner edge
semi-major axis of 1\farcs 27 (180 AU); outer edge semi-major axis of 
1\farcs 73 (240 AU); and a ring inclination 37$^{\circ}$ from face-on.  
The center of the projected ellipses is offset 0\farcs 26 S of the binary 
center of mass.
In the case of a completely flat, elliptical ring this would be a physical
offset, and an intrinsic ring eccentricity of 0.21 would be required.
However, it is much more likely that the ring has a significant vertical 
thickness that is 10-20\% of its radius.  In this 
case, even an intrinsically circular ring can exhibit an offset in the 
center of its scattered light from its central star(s).  In GG Tau
the apparent offset is in the direction expected for the latter case (GDS99).
Therefore, the offset of the ring center from the binary is not a compelling 
reason to ascribe any intrinsic eccentricity to the circumbinary ring.
Uncertainty in the precise value of the ring thickness prevents us from
drawing any conclusions about the ring eccentricity from simple ellipse 
fitting.

The azimuthal distribution of light in the ring is somewhat of a mystery.
CO kinematics indicate that the near side is located at
PA=7$^{\circ}$.  In a symmetric, circular ring with forward-scattering dust
grains, the brightest azimuth would be expected to align
exactly at this position angle, but in GG Tau it does not.  Instead,
the azimuth of peak brightness is offset $\sim13^{\circ}$ E from this, as
determined by visually fitting a curve to the brightest section of the
azimuthal profile.  A curve fit to a larger section (PA=$-80^{\circ}$ to 
$90^{\circ}$) peaks at about 7$^{\circ}$, as expected, indicating that
the offset is a localized phenomenon.   The effect of binary illumination 
should not shift the azimuth of peak brightness 
significantly from the expected value, as discussed below.  Shadowing by
a non-coplanar disk or perhaps the material seen near the secondary might
result in the offset.  Another explanation might be an intrinsic ring 
eccentricity.  In this case, the azimuth of peak brightness would be offset 
away from the PA of the near edge and toward the PA of ring periapse.
While this possibility cannot be discounted, other evidence for ring
eccentricity would be needed to establish it.

Comparison of scattered light models with optical HST disk images has
proven to be a valuable way to constrain the disk density distributions
and dust properties (Burrows et al. 1996; Stapelfeldt et al. 1998).
Unfortunately it was not practical for us to determine an optimal model
for the GG Tauri ring.  The key problem is uncertainty in the relative
contributions of each star to the ring illumination.  At one extreme
is the possibility that one star dominates the ring illumination,
as suggested by the fact that the primary appears $>$ 1.5 mag brighter
than the secondary at optical wavelengths.  At the other extreme is the
possibility that each star makes a roughly equal contribution to the ring
illumination, as suggested by the fact that the components have comparable
intrinsic brightnesses after correction is made for extinction (White et
al. 1999).  Given this range of uncertainty, we did not attempt to identify
a fully optimized scattered light model for GG Tauri.  Instead, we chose
only to validate the disk density distribution determined by GDS99 as a
consistency check with the WFPC2 images.  To do this, we used the PINBALL
Monte Carlo multiple scattering code (Watson \& Henney 2001) to calculate 
model ring images at $\lambda$ = 0.8 $\mu$m.  An inclined circular ring 
was illuminated by identical binary point sources whose deprojected
positions were consistent with the parameters of GG Tau.  

\begin{figure}
\plotone{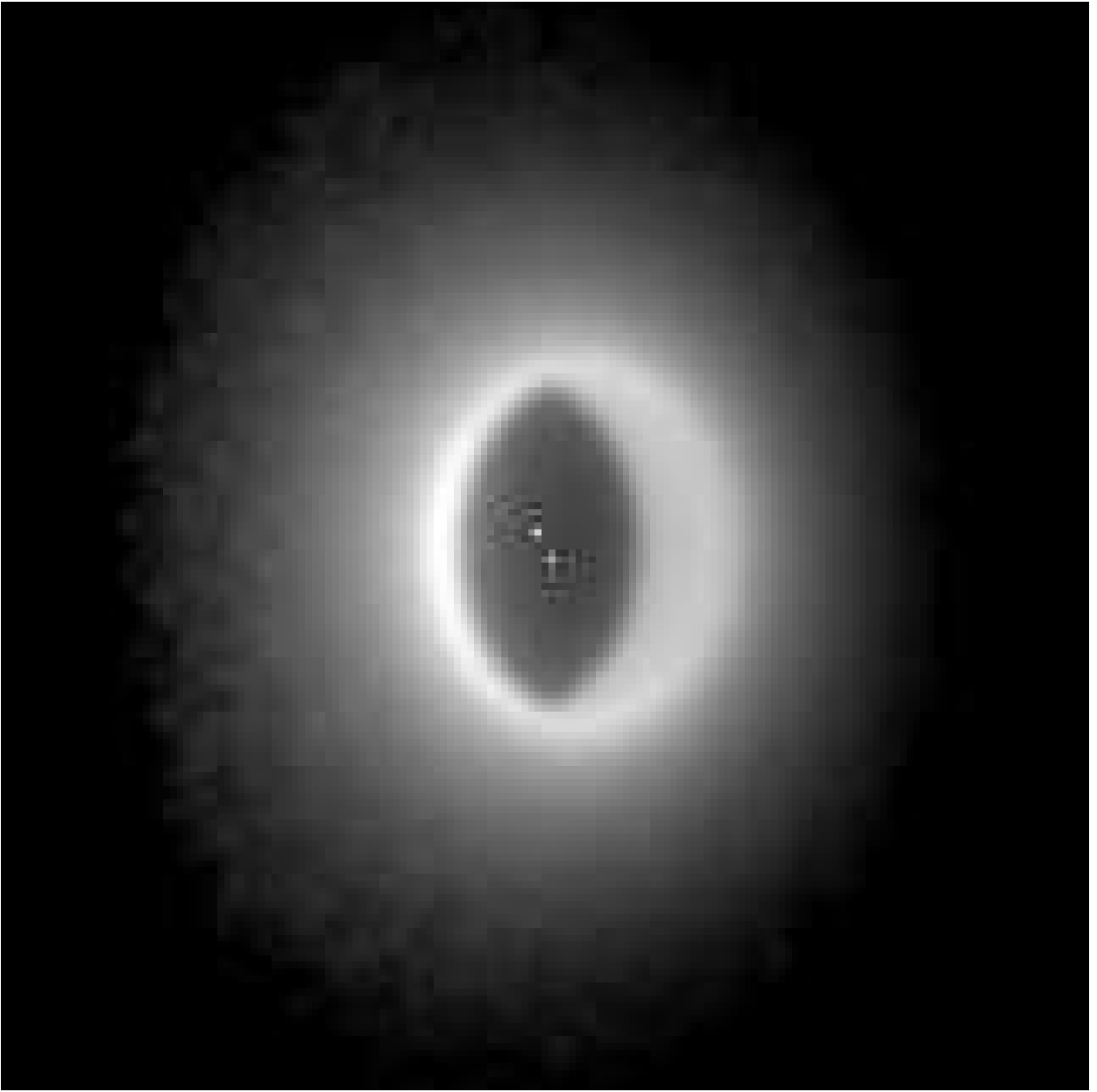}
\caption{
Monte Carlo multiple scattering model of the GG Tauri circumbinary disk, 
displayed with a logarithmic stretch and with a similar orientation
as the ring shown in Figures 1 and 2.  Illumination from equally bright binary 
components was assumed (no circumstellar disks).   The model parameters are: 
inner radius $=$ 180 AU; outer radius $=$ 800 AU; scale height = 40 AU at 
300 AU; disk mass $=$ 0.12 M$_{\odot}$; scale height scales with radius 
as $r^{1.05}$; midplane volume density scales with radius as $r^{-2.75}$; 
dust opacity 120 cm$^2$ gm$^{-1}$; scattering phase function asymmetry 
parameter $g$ = 0.65; inclination is 37$^{\circ}$; and 190 million photons 
traced.  Dots mark the positions of the stars.  The disk model (but not 
the stars) was convolved with a Tiny Tim model PSF for filter F814W.}
\end{figure}

In these simulations, the GDS99 density law provides a good  match to the
morphology of the WFPC2 image.  This model is shown in Figure 5.
Through trial and error, we found that a Henyey-Greenstein phase function
asymmetry parameter {\it g}=0.65 provided the best match to the observed
brightness ratio between the front/back sides of the ring.  This value
is in excellent agreement with determinations made in the disks of HH 30
(Burrows et al. 1996) and HK Tau/c (Stapelfeldt et al. 1998).  Finally,
our simulation shows that even a continuous disk extending from 
180 AU $< r <$ 800 AU is dominated by a bright inner ring of light like
that seen in GG Tau.  Both the WFPC2 data and the scattered light model
show that on disk's N side, the surface brightness drops by an order of
magnitude just 0\farcs 3 outside the radial peak emission of the bright ring.
No density discontinuity or enhanced shadowing is needed to account for 
the different outer radii measured in scattered light and millimeterwave 
CO emission; such a sharp drop in surface brightness with radius is
expected, even for a continuous disk.

For the assumed case of equal binary illumination, the distribution of light 
on the ring appears nearly the same as for the case where the ring is 
illuminated by a single central point source.  This result can be readily 
understood analytically.  The peak-to-valley deviation of the disk azimuthal
brightness profile in a
coplanar binary model with both stars of equal brightness compared to a 
single-star model is ${\cal R}= [(R+d)^2 + (R-d)^2] / 2 R^2$, where $R$ is the
disk inner radius and $d$ is half the binary separation.  Defining 
$x\equiv d/R$, then  ${\cal R}= [(1+x)^2 + (1-x)^2] / 2 = 1 + x^2$.  
For the case of GG Tau, $R= $ 180 AU at the ring inner edge, and 
$d= $22 AU (deprojected): thus, the peak-to-valley variation is only 3\% 
around the ring.  The peak-to-valley deviation of disk brightness 
in a coplanar binary model with one dominant star compared to a single-star 
model is  ${\cal R}= (R+d)^2 / R^2$, which reduces to $1 + 2x$.  This would 
produce a 40\% peak-to-valley effect. Thus, the surface brightness of the 
disk will only differ significantly if one star (as seen by the disk)
is much brighter than the other.

\subsection{Ring photometry}

The reflected light from the ring appears redder than the direct starlight
in the WFPC2 optical images, continuing the trend first identified in the
near-infrared by R96.  In an optically thick disk like that of GG Tau,
the color of reflected starlight should be very similar to the color of
the source(s) which illuminate it.  The color difference between the
disk light and the direct starlight (V-I= 0.8 mag) therefore requires
that light seen by the disk be reddened by an additional $A_V \ge 1.2$
mag with respect to the direct light seen by an earthbound observer
(assuming a standard extinction law).  Wood, Crosas, \& Ghez (1999) have 
shown that such an extinction can be readily produced by circumstellar disks 
surrounding one or both of the central binary components. 

We searched for color gradients in the ring using a smoothed F555W/F814W
ratio image.  Two possible mechanisms might produce observable color
gradients in the ring: (1) Any regions with enhanced shadowing would be
expected to appear redder than the rest of the ring; and (2) the expected
wavelength dependence in the dust phase function should cause the near (N)
side of the ring to appear more blue, and the far (S) side to appear redder,
than the ring as a whole.  Unfortunately we were not able to identify
any color gradients in the ring  because of the low signal-to-noise and large 
PSF subtraction residuals in our F555W image.

\subsection{Ring substructure}

The WFPC2 and NICMOS images indicate that the GG Tau ring is essentially
smooth and not clumpy as the adaptive optics (AO) images of R96 would
indicate.  Also, the spokes seen in the AO images, some of which are
significantly brighter than the ring, are not apparent.  These clumps
and spokes may be instrumental artifacts.  AO systems can produce  
artificial structures around bright stars, especially when deconvolution 
techniques are employed.  The brightest
spoke in the AO images, which connects the stars to the northern, bright
edge of the ring, may actually be a merge of the flux from the ring and
the material that we see adjacent to GG Tau {\it Ab}.

The gap in the ring surface brightness at PA=268$^{\circ}$ is reminiscent of
the azimuthal density structure predicted for the case of a low mass
companion embedded within a circumstellar debris disk (Liou \& Zook
1999).  Without dynamical forcing, any sharp azimuthal density
structure corresponding to the gap would be disrupted by Keplerian shear
in a few orbital timescales.  Also, there are no indications of a gap
in the millimeter data.  An alternative possibility is that the
gap represents some kind of shadow cast onto the ring by intervening,
material, unseen in our images, lying between it and the illuminating stars.  
The shadow would
have to block the illumination of the brightest component of the binary,
otherwise it would not be possible to produce the observed brightness
deficit of $>$1 mag in the gap.  Furthermore, it would be necessary
to assume that one star dominates the illumination of the outer ring -
otherwise the gap would be filled in by the illumination of the other
binary component.  With these caveats, there are three possibilities for
the nature of the shadowing material, none of which is fully satisfactory.
The first possibility is that of a small circumstellar disk at the
brightest binary component, extending along PA=268$^{\circ}$, and inclined by a
large angle to the plane of the outer ring.  Illumination of the outer
ring along the midplane of this circumstellar disk would be suppressed,
leading to a darkened gap along the line of intersection between the
outer ring and inner disk planes.  While this model would do a nice job
explaining the observed gap, it also predicts that a second, diametrically
opposed gap should be present.  This is not observed.  A second, more
ad-hoc possibility is to postulate a circumstellar disk coplanar with the
outer ring, and with a large azimuthal density enhancement that shadows
the outer ring to produce the gap.  Finally, if mass transfer from the
outer ring to the inner circumstellar disks were actually taking place,
it is possible to imagine a dense clump in the accretion stream that
might shadow the outer ring.  With none of these explanations finding
immediate support in the available data, for now the gap's origin must
remain a mystery.

\subsection{Circumsecondary material}

The WFPC2 F814W observations show an arc of nebulosity located near
the secondary star and well within the circumbinary ring.  Given the
strong artifacts present in the image just 0\farcs 3 from the secondary, it
is unclear whether this nebulosity is part of a continuous distribution
of material extending outward from the secondary, or a localized clump.
If continuous, then a natural explanation would be a circumstellar
disk surrounding GG Tau {\it Ab}.  The infrared excess and strongly localized
extinction indicate a significant amount of dust near this star.
A circumstellar disk coplanar with the ring would appear brightest on
its N side because of enhanced forward scattering, exactly where the arc
feature is found.  However, the feature's location (0\farcs 3 outward from
the secondary) implies an extent considerably larger than expected for
a disk, whose maximum outer radius should be 3-5 times smaller than binary
separation (i.e., $\le$ 0.1 arcsec) due to tidal truncation.  While this 
feature is small, it is still much larger than the core of the WFPC2 PSF - 
so even such a barely resolved disk could not plausibly extend to this radius 
when convolved with the PSF.

The location of this reflecting material suggests another possible
explanation, that of material moving in a two body potential.  The
binary's Lagrange stability point L2 is located on the line passing
through the two stars, and on the opposite side of the secondary from
the primary - near the location of this nebular arc.  In a binary with the
relative stellar masses inferred for GG Tau, this stability point would be located
opposite the primary and offset from the secondary by 67\% of the
primary-secondary separation (Taff 1985).  Since the material
near the secondary is displaced from the star by more than this expected
value (0\farcs 3 versus $\sim$0\farcs 17), it appears that this is also 
not a satisfactory explanation for the nature of the nebular arc.

\subsection{Monitoring possible time variations}

Multiple-epoch observations of circumstellar disks (Watson et al. 2000;
Wood et al. 2000)
indicate that shadowing from variations in the dust density distribution 
in the inner disk, or beaming from accretion hot spots at the stellar surface, 
can significantly alter the light distribution in the outer disk, with
timescales of hours to years.  In the case of GG Tau, where shadowing by
circumstellar disks may be necessary to explain the properties of the 
circumbinary ring, there may likewise be time-variable illumination effects.
Orbital timescales in the circumstellar disks are on the order of 30 years,
so the motion of any shadow patterns might be discernible in just
a fraction of this period.  In addition, a section of the ring may
also reflect variations in the brightness of the star closest to it,
skewing the front/backside brightness ratio or the azimuthal ring profile.
Possible orbital motion of the dust in the arc near the secondary would
also be worth investigating, as it would have a timescale on the order of
a few decades.
It would be interesting to monitor any such changes in the GG Tau system
on long and short timescales using HST, adaptive optics, or in the future 
with the Next Generation Space Telescope. 

\section{Conclusions}

HST/WFPC2 V and I band images of GG Tau resolve the binary star and reveal the 
scattered light from the circumbinary dust ring that was previously detected 
in the near-IR.  The images confirm that the northern, nearer side of the ring 
is brightest, as is expected from forward scattering by dust grains.  The 
nearside/farside brightness ratio is 6.9 in the I band.  This effect 
has been reproduced using three-dimensional scattering models of the ring.
The V-I$_c$ color of the ring relative to the binary colors indicates that some 
amount of extinction must be introduced by material between the stars, as first 
pointed out by R96.  The new images show a clump of reflecting material near the 
secondary star, also seen in the Silber et al. (2000) HST images (though apparently
not noticed by them), that could 
possibly create shadows on, or at least introduce partial extinction to, the ring.
Clear evidence for shadowing is provided by a narrow gap in the ring, which has 
also been seen in HST/NICMOS and adaptive optics images.  However, we cannot 
determine what may be causing this shadow, nor why it is only seen on one side if 
it is created by a circumstellar disk.  Other brightness asymmetries, such as the 
differences in the eastern and western ring radial profiles, and the offset of the
ring peak brightness from the closest edge, might be explained by shadowing
or possibly by warping of the ring.  The observations highlight the fact that the
poorly understood interaction of light from the binary with ring material and 
possible circumstellar disks is the primary limitation to modelling the dust 
distribution as derived from scattered light images.  

We find no evidence in our images for the ``spokes'' or ring clumpiness seen 
in the adaptive optics observations of R96.  We believe that these may be AO
artifacts.  However, we do see some indication that the interior of the ring 
may not be completely clear, as there is a generally uniform distribution of
light within the ring that is above the background level.  More observations 
are required to verify whether this is reflective material or simply a PSF
subtraction residual. 
 
Higher resolution imagery, which may provide details of the circumstellar 
disks, should substantially improve our understanding of GG Tau.  Ground-based
adaptive optics imaging will help, but as demonstrated by the possible 
artifacts in the R96 images, any details will need to be confirmed by other 
systems.  In the near future, the Advanced Camera for Surveys (ACS) on HST 
will provide high resolution (0\farcs 025 arcsec pixel$^{-1}$) visible-
wavelength frames with a non-field-dependent PSF, which should improve PSF 
subtraction results compared to WFPC2.  GG Tau has been included in the ACS 
Science Team's program.  Further into the future, observations with ALMA 
(Atacama Large Millimeter Array) should provide high resolution (0\farcs 01 - 
0\farcs 1) maps of disk emission, avoiding the complications introduced by 
shadowing and extinction in scattered light images.

\section{Acknowledgements}
The GG Tauri data were taken as part of the WFPC2 Science Team's program.
Observations were made with the NASA/ESA Hubble Space Telescope, obtained 
from the Data Archive at the Space Telescope Science Institute, which
is operated by the Association of Universities for Research in Astronomy, 
Inc., under NASA contract NAS 5-26555.

\end{document}